\documentclass{aa}  

\usepackage{graphicx}
\usepackage{txfonts}
\usepackage{hyperref}  
\hypersetup{colorlinks=true,linkcolor=[rgb]{1.,0.2,0.2},citecolor=[rgb]{0.1,0.4,1.},filecolor=[rgb]{0.7,0.2,0.2},urlcolor=[rgb]{0.7,0.2,0.2}}

\newcommand{\ha}{\ifmmode {\rm H}\alpha \else H$\alpha$\fi}
\newcommand{\hb}{\ifmmode {\rm H}\beta \else H$\beta$\fi}
\newcommand{\lya}{\ifmmode {\rm Ly}\alpha \else Ly$\alpha$\fi}
\newcommand{\pg}{\ifmmode {\rm P}\gamma \else Pa$\gamma$\fi}
\newcommand{\lyb}{\ifmmode {\rm Ly}\beta \else Ly$\beta$\fi}
\newcommand{\lyg}{\ifmmode {\rm Ly}\gamma \else Ly$\gamma$\fi}

\newcommand{\ciiidoub}{\textrm{C}\textsc{iii}]\ensuremath{\lambda\lambda1907,1909}}

\newcommand{\siiiuvem}{[\textrm{Si}\textsc{iii}]\ensuremath{\lambda1883,1892}}

\newcommand{\nv}{\textrm{N}\textsc{v}\ensuremath{\lambda1240}}

\newcommand{\civmed}{\textrm{C}\textsc{iv}\ensuremath{\lambda 1550}}
\newcommand{\heii}{\textrm{He}\textsc{ii}\ensuremath{\lambda1640}}
\newcommand{\niv}{\textrm{N}\textsc{iv}\ensuremath{\lambda1720}}
\newcommand{\nivblue}{\textrm{N}\textsc{iv}\ensuremath{\lambda1486}}\newcommand{\niiiuv}{\textrm{N}\textsc{iii}]\ensuremath{\lambda1750}}
\newcommand{\ov}{\textrm{O}\textsc{v}\ensuremath{\lambda1371}}
\newcommand{\oiiiuv}{\textrm{O}\textsc{iii}]\ensuremath{\lambda1661,1666}}

\newcommand{\flyc}{\ifmmode  \mathrm{f}_\mathrm{esc}\mathrm{(LyC)} \else $\mathrm{f}_\mathrm{esc}\mathrm{(LyC)}$\fi}

\def\kms{km s$^{-1}$}

\def\ergs{\ifmmode \mathrm{erg\hspace{1mm}s}^{-1} \else erg s$^{-1}$\fi}

\def\micron{\ifmmode \mu\mathrm{m} \else $\mu$m\fi}
\def\msun{\ifmmode \mathrm{M}_{\odot} \else M$_{\odot}$\fi}
\def\msunyr{\ifmmode \mathrm{M}_{\odot} \hspace{1mm}{\rm yr}^{-1} \else $\mathrm{M}_{\odot}$ yr$^{-1}$\fi}
\def\zsun{\ifmmode Z_{\odot} \else Z$_{\odot}$\fi}
\def\lsun{\ifmmode L_{\odot} \else L$_{\odot}$\fi}
\def\mstar{\ifmmode \mathrm{M}_{\star} \else M$_{\star}$\fi}

\begin{document}

\titlerunning{Very massive spatially segregated stars at z=2.4}
\title{
Clues on the  presence and segregation of very massive stars in the Sunburst Lyman-continuum cluster at z=2.37\thanks{Based on observations collected at the European Southern Observatory for Astronomical research in the Southern Hemisphere under ESO programmes DDT MUSE program ID 107.22SK.001 (PI E. Vanzella), X-shooter program ID 0103.A-0688 (PI E. Vanzella) and DDT MUSE program ID 297.A-5012(A) (PI Aghanim). }}

\authorrunning{U. Mestric et al.}
\author{ 
U. Me\v{s}tri\'{c} \inst{\ref{inafbo}} \fnmsep\thanks{E-mail: \href{mailto:uros.mestric@inaf.it}{uros.mestric@inaf.it}},
E. Vanzella \inst{\ref{inafbo}},
A. Upadhyaya \inst{\ref{unige}},
F. Martins \inst{\ref{lupm}},
R. Marques-Chaves \inst{\ref{unige}},
D. Schaerer \inst{\ref{unige},\ref{CNRS}},\\
J. Guibert \inst{\ref{unige}},
A. Zanella \inst{\ref{inafpd}},
C. Grillo \inst{\ref{unimi},\ref{iasf}},
P. Rosati \inst{\ref{unife}},
F. Calura \inst{\ref{inafbo}},
G.B. Caminha \inst{\ref{tum},\ref{MPI}},
A. Bolamperti \inst{\ref{inafpd},\ref{unipd},\ref{ESO}},\\
M. Meneghetti \inst{\ref{inafbo}},
P. Bergamini \inst{\ref{inafbo},\ref{unimi}},
A. Mercurio \inst{\ref{dfc},\ref{inafna}},
M. Nonino \inst{\ref{inafts}},
R. Pascale \inst{\ref{inafbo}}
}

\institute{
INAF -- OAS, Osservatorio di Astrofisica e Scienza dello Spazio di Bologna, via Gobetti 93/3, I-40129 Bologna, Italy \label{inafbo} 
\and
Geneva Observatory, Department of Astronomy, University of Geneva, Chemin Pegasi 51, CH-1290 Versoix, Switzerland \label{unige}
\and
LUPM, Universit\'e de Montpellier, CNRS, Place Eug\`ene Bataillon, F-34095 Montpellier, France
\label{lupm}
\and
CNRS, IRAP, 14 Avenue E. Belin, 31400 Toulouse, France \label{CNRS}
\and
INAF -- Osservatorio Astronomico di Padova, Vicolo dell'Osservatorio 5, 35122, Padova, Italy \label{inafpd}
\and
Dipartimento di Fisica, Universit\`a  degli Studi di Milano, via Celoria 16, I-20133 Milano, Italy \label{unimi}
\and
INAF -- IASF Milano, via A. Corti 12, I-20133 Milano, Italy \label{iasf}
\and
Dipartimento di Fisica e Scienze della Terra, Universit\`a degli Studi di Ferrara, via Saragat 1, I-44122 Ferrara, Italy \label{unife}
\and
Technical University of Munich, TUM School of Natural Sciences, 
Department of Physics, James-Franck-Str 1, 85748 Garching, Germany 
\label{tum}
\and
Max-Planck-Institut f\"ur Astrophysik, Karl-Schwarzschild-Str. 1, D-85748 Garching, Germany \label{MPI}
\and
Dipartimento di Fisica e Astronomia, Università degli Studi di Padova, Vicolo dell’Osservatorio 3, I-35122 Padova, Italy \label{unipd}
\and
European Southern Observatory, Karl-Schwarzschild-Strasse 2, D-85748 Garching bei M\"unchen, Germany \label{ESO}
\and
Dipartimento di Fisica “E.R. Caianiello”, Universit\`a Degli Studi di Salerno, Via Giovanni Paolo II, I–84084 Fisciano (SA), Italy \label{dfc}
\and
INAF -- INAF - Osservatorio Astronomico di Capodimonte, Via Moiariello 16, 80131 Napoli, Italy 
\label{inafna}
\and
INAF -- Osservatorio Astronomico di Trieste, via G. B. Tiepolo 11, I-34143, Trieste, Italy \label{inafts}
}

\date{} 

 
\abstract
{We report the identification of very massive stars (VMS; mass $> 100$\,\msun) that may be segregated in the center of the young massive star cluster at $z$=2.37 hosted in the lensed galaxy called {\tt Sunburst} galaxy. This result is based on two pieces of evidence: (1) VLT/MUSE spectra of several multiple images of the same star cluster show key spectral signatures of VMS, such as the \heii\ broad emission, \nivblue\ emission, and an \niv\ P-Cygni profile. In particular, \heii\ is broad ($\sim1610\pm300$ \kms), with an equivalent width of 3\AA,\ and asymmetric profile. These features require an extremely young ($\sim2.5$ Myr) stellar population component in which the masses of the stars exceed 100~\msun. When a Salpeter initial mass function and BPASS models for normal massive stars are assumed, the observed spectral features require $\sim$400 VMS. (2) The same star cluster is detected at a signal-to-noise ratio of~$\sim100$ in the Lyman continuum domain ($\lambda < 900$\AA).  
The Lyman continuum emission emerges from a region with a radius that is at least twice smaller than what is observed at 1700\AA~(independently of  magnification) and is located in the center of the cluster. After delensing, the effective radii in absolute scales are R$_{\tt eff}[{\tt LyC}]\sim4.7 \pm 1.5$ pc and R$_{\tt eff}[1700]= 7.8 \pm 1.4$ pc.
The Lyman continuum radiation is mainly produced by hot and massive stars, which implies that their spatial distribution (including that of VMS) is preferentially more confined in the central parts of the cluster. 
Approximately 400 VMS hosted by a cluster of $\sim 10^7$ \msun\  produce $\sim$15\% of the escaping Lyman continuum photons, and the remaining photons are produced by other massive early-type stars.
}

   \maketitle
%

\section{Introduction}

For many years, the existence and occurrence of very massive stars (VMS) was mostly associated with the early Universe and metal-free environments in the context of the so-called Population III stars \citep[e.g.,][]{Abel2002}.
VMS are short-lived stars that are $\sim$ 2 -- 3 Myr old \citep[e.g.,][]{Yusof2013}, have a mass $M>100$\,\msun\ \citep{Vink2015}, and predominantly populate the central regions of young massive star clusters \citep[within the core radius $r_c \sim 0.1-0.2$ pc;][]{zwart2010}.
Studies of VMS in Milky Way star clusters are limited to only few targets because their lifetimes are so short. Example are the Arches cluster \citep{Martins2008} or NGC3603 \citep{Crowther2010}.
Individual VMS have been investigated in the local Universe with high spatial resolution with the Hubble Space Telescope \citep[HST;][]{Cignoni2015, Crowther2016, calzetti15_ngc5253, smith16, Smith2020, Brands2022}.
Very massive stars with masses higher than 100 $\msun$ are recognized as objects that strongly affect the evolution of early galaxies in their chemical enrichment and star formation through feedback \citep[e.g.,][]{Goswami2021}.
It is therefore essential to extend the upper masses beyond 100 \msun\ of the current population synthesis models to investigate and understand young massive star clusters and VMS at different redshifts \citep{smith16,Crowther2016}.

Despite some progress, the maximum stellar mass that can be attained and the conditions determining the presence of VMS largely remain unknown.
Recent observations of local star clusters reported initial stellar masses up to $\sim270$ $\msun$ \citep{Brands2022} in the star cluster R136, with a cluster age of $\sim$1.5 Myr \citep{Crowther2016}. 
Furthermore, observations of young stellar clusters have revealed peculiar spectroscopic features such as unusually strong broad \heii\ emission (with full width at half maximum (FWHM)~$>1000$~\kms), which suggests the presence of VMS in these objects \citep[e.g.,][]{Wofford2014, Crowther2016, Senchyna2021}.

Alongside with the observations, different models are trying to predict and trace the evolution of the various spectroscopic features characteristic of VMS through different ages and masses  \citep[e.g.,][]{Koehler2015, Graefener2021}.
For example, 
\citet{martins2022} have generated new evolutionary models and synthetic spectra of stars with initial masses in the range 150 -- 400 $\msun$, taking the specific winds of VMS into account that are stronger from a simple extrapolation of OB-type winds at high luminosity. 
The resulting models predict specific features in the UV and optical part of the spectra that are characteristic signatures of VMS.
The most robust ultraviolet spectral features associated with VMS are \nivblue, broad \heii\ \citep[and narrow; see][]{Grafener2015} emission, and the \niv\ P-Cygni profile \citep{martins2022}.
Equivalent widths of these lines are expected to span the interval of $0.1-7$~\AA\ in the rest frame. 
Spectra with a high signal-to-noise ratio (S/N) with well-detected continuum are therefore required to identify them, as was shown by \citet{Crowther2016} in the R136 stellar cluster in the local Universe, for example. 
At cosmological distance, strong gravitational lensing is necessary to detect these faint spectral features, which allows us to further gain in spatial resolution at a scale of tens of parsec and depth \citep[see also][]{vanz16_id11, johnson17, rigby17, rigby18_I, rigby18_II, vanz_paving, vanz_mdlf, UrosM22, vanz_glass_protoGC_2022}.
 
We present convincing spectroscopic evidence for VMS in a stellar cluster at cosmological distance for the first time ($z$=2.37; \citealt{vanz_sunburst,vanz22_CFE}).
The host galaxy is called {\tt Sunburst} \citep[][]{rivera19, rivera17}, and Lyman continuum (LyC) radiation is detected from the same clumpy regions in which the VMS were detected.
These massive stars in the center of the stellar cluster are significant producers of LyC radiation, through which a porous interstellar medium (ISM) is created that enables LyC escape. 
For the purpose of this work, we performed a comprehensive analysis of deep Very Large Telescope (VLT) / Multi Unit Spectroscopic Explorer (MUSE), X-shooter, and synthetic spectra with the aim to confirm the presence of the VMS. 
Additionally, we investigated whether VMS are segregated by modeling the morphology of the young massive star cluster (YMC) that is hosted in the {\tt Sunburst} galaxy. 

The paper is organized as follows.
In Section \ref{sec:Sunburst} we briefly describe the {\tt Sunburst} galaxy and the available observational data.
In Section \ref{sec:VMS} we analyze the spectral signatures of very massive stars using MUSE/IFU and X-shooter observations in combination with the latest evolutionary models and synthetic spectra. In Section \ref{sec:morpho} we discuss the morphological properties of the YMC (called source 5.1) and the possible segregation of the (very) massive stars in its central parts. We present our conclusion in Section \ref{sec:Summ}. 

We assume a flat cosmology with $\Omega_{M}$= 0.3,
$\Omega_{\Lambda}$= 0.7 and $H_{0} = 70$ km s$^{-1}$ Mpc$^{-1}$. Within this model, one arcsec at $z=2.37$ corresponds to a projected physical scale of 8200 parsec. All magnitudes are given in the AB system.

\section{Lensed galaxy {\tt Sunburst} }
\label{sec:Sunburst}

\begin{figure}
        \centering
        \includegraphics[width=\linewidth]{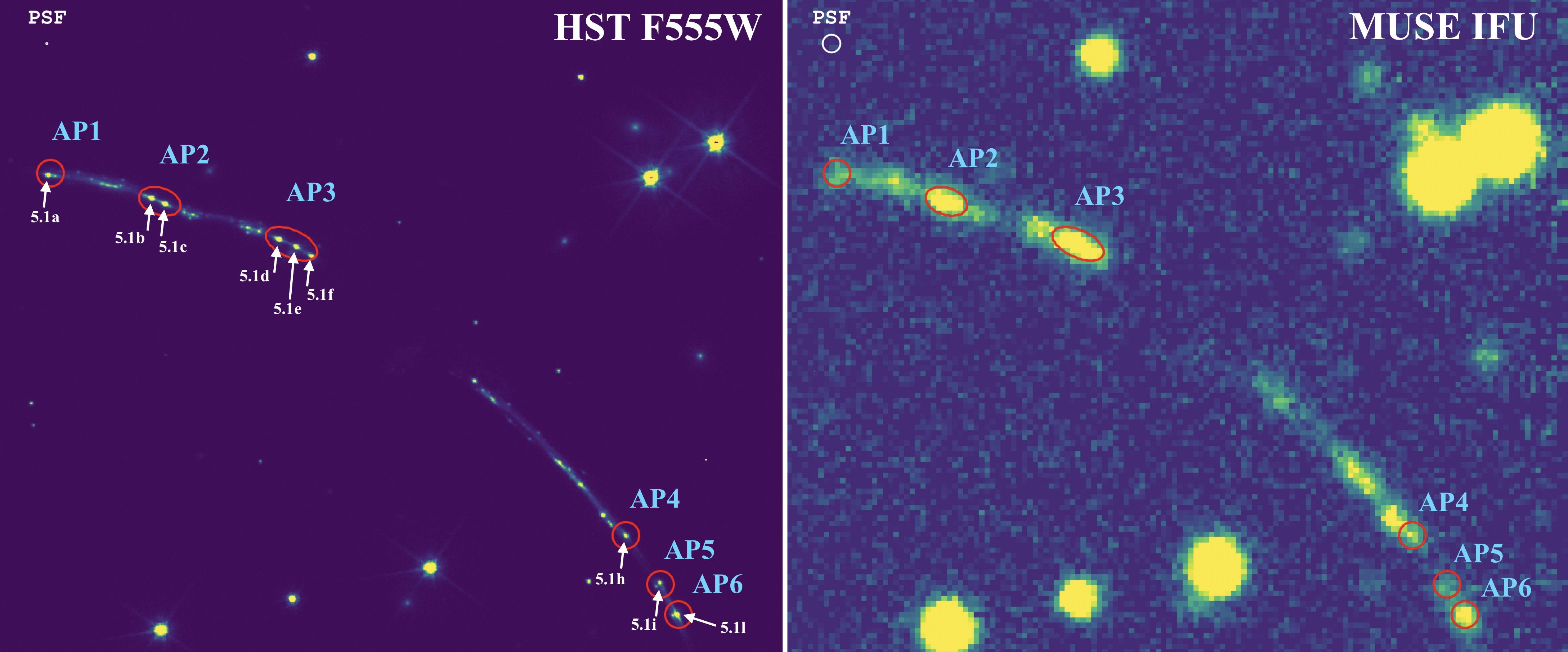}
        \caption{Sunburst bright arcs with a visible young massive stellar cluster. \textbf{Left:} HST F555W band image, showing the six aperture positions from which a MUSE 1D spectrum is extracted (red contours). White arrows point to the multiple images of the young stellar cluster. \textbf{Right:} MUSE IFU image at $\sim 1800\AA$ of the same region as on the left. The same apertures are plotted in red.} 
        \label{spec_ap}
\end{figure}

The {\tt Sunburst} galaxy is a galaxy at $z$=2.37. It is strongly lensed by the Planck cluster PSZ1 G311.65-18.48 at $z$=0.44, as initially reported by \cite{dahle16}. 
The strong gravitational lensing effect deflects the light from the background high-$z$ {\tt Sunburst} galaxy into  four bright arcs. 
These bright arcs harbor at least 13 star-forming knots that likely are stellar clusters.
There are more than 50 multiple images of this system \citep[][]{Pignataro2021}, whose physical properties are studied in detail in \cite{vanz22_CFE}.
One of the 13 young stellar clusters has been identified 12 times (called source 5.1) and is the subject of this work (see Figure \ref{spec_ap}).
The multi-peaked \lya\ emission of source 5.1 is consistent with an optically thin medium and LyC leakage along the line of sight \citep{rivera17}.
Furthermore, the detection of LyC radiation emerging from the 12 detected multiple images of the young massive star cluster in source 5.1 was confirmed by HST multiband observations \citep{rivera19}.
Additional analyses of the 12 LyC multiple images of source 5.1 have revealed that the star cluster is younger than 3 Myr and has a stellar metallicity  of $0.5 Z_{\sun}$ \citep{chisholm19}, with a physical size of $\simeq 10$ pc and a stellar (and dynamical) mass value of $\simeq 10^{7}$ \msun\  \citep{vanz22_CFE}. 

The {\tt Sunburst} was observed with the HST, providing multiband photometry in the F275W, F410M, F555W, F606W, F814W, F098M, F105W, F140W, and F160W filters under the programs 15101 (PI Dahle), 15949 (PI Gladders), and 15377 (PI Bayliss). 
{\tt Sunburst} has also been targeted with ground-based high-resolution ($R\sim 5000-9000$) VLT/X-shooter spectroscopy covering the spectral range 3000-22000\AA\ in three main arms, UVB, VIS, and NIR. The observational strategy and the data reduction procedures applied to HST imaging and VLT/X-shooter spectroscopy  have been presented in \cite{vanz_fluoresc., vanz22_CFE}.
VLT/MUSE integral field spectroscopy at a resolution $R=3000$ and covering the spectral range 4800-9400\AA\ was obtained during 2016 (1h integration, DDT; PI: Aghanim) and 2021 (1h integration; PI: Vanzella) in the wide-field mode configuration. The final datacube that combines the two hours and the data reduction is described in \citet{vanz22_CFE}.
We also presented a first version of the lens model in \cite{Pignataro2021} based on the 62 spectroscopically confirmed multiple images in the redshift range $1<z<3.5$
(see also \citealt{Sharon2022, Diego2022}). A revised lens model will be computed when the new VLT/MUSE observations (7h integration) that are planned for 2023 will be performed (prog. 110.249D.001; PI: Vanzella).

Here we focus on the $\simeq 3$ Myr old, UV-bright, and LyC source with M$_{\rm UV}=-18.6$ (1700\AA\ magnitude and ultraviolet slope $\beta=-1.71\pm0.01$, F$_{\lambda} \sim \lambda^{\beta}$), massive (M~$\sim10^{7}$~\msun) star cluster 5.1, which has high magnification values ($\mu\sim10-70$ over 12 multiple images; \citealt{Pignataro2021, vanz22_CFE}). 
In the following, we perform a new analysis focusing on the nature of the ionizing source (Sect.~\ref{sec:VMS}) and its morphology (Sect.~\ref{sec:morpho}).

\begin{figure*}
        \centering
        \includegraphics[width=\linewidth]{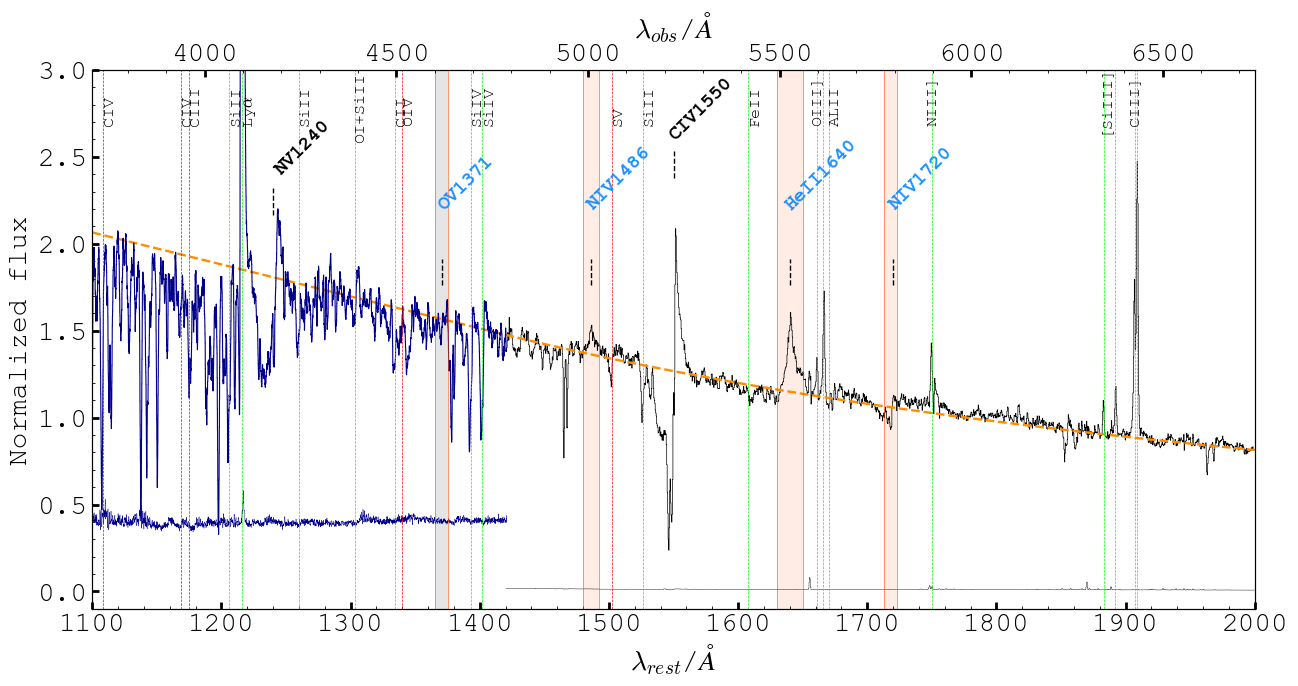}
        \caption{MUSE IFU spectrum of the young massive star cluster 5.1 extracted from six apertures (thin black line) and the X-shooter long-slit spectrum of 5.1l knot (bold blue line). The key confirmed features indicating the presence of VMS in the stellar cluster are marked with shaded light red strips, and the dark orange line shows the best-fit F$_{\lambda} \sim \lambda^{\beta}$, with $\beta = -1.71$. 
        The shaded gray strip indicates the (absence of) \ov\ line, which usually is an indicator of VMS too. Prominent P-Cygni of \nv\ and the strong emission part of the \civmed\  indicate the young age of the stellar cluster (bold black markers). In the bottom panel, the $1\sigma$ errors of both spectra are shown. Other detected interstellar features and stellar features are marked with dashed red and green lines, respectively. }
        \label{1Dspec}
\end{figure*}

\section{Spectral signatures of very massive stars in the {\tt Sunburst} star cluster at z=2.37}
\label{sec:VMS}

We investigate the UV and optical spectroscopic properties of the young stellar cluster 5.1. 
The VLT/MUSE one-dimensional spectra were extracted from six apertures enclosing nine multiple images of source 5.1 (shown in Fig.~\ref{spec_ap}) and 
were subsequently combined to produce a continuum-detected high S/N~($>60$) weighted-average spectrum (Figure~\ref{1Dspec}). 
The stacked spectrum shown in Figure~\ref{1Dspec} is 
equivalent to an integration time of $(2 \times 9)\times 30^{2} > 16,000$  hours without lensing amplification when the minimum amplification in the 9 multiple images is adopted ($\mu=30$).

\subsection{Observed VMS features with VLT MUSE and X-Shooter}

The very high S/N MUSE spectrum (Fig.~\ref{1Dspec}) allows us to identify several emission and absorption lines. 
They include the nebular emission lines associated with the interstellar medium of the galaxy, such as \oiiiuv, \niiiuv, \siiiuvem, and \ciiidoub. 
The well-detected continuum allows us to investigate faint line emissions (of a fraction of an \AA\ rest-frame equivalent width), and to sample the details of the line profiles that are otherwise not accessible without lensing amplification. 
In particular, faint \nivblue\ emission, the evident P-Cygni profile of the \civmed, the prominent broad and asymmetric \heii\ line profile, and the P-Cygni signature of \niv\ clearly stand out.
All these lines are associated with young, hot, and (very) massive stars.

We report the detection of \heii\ emission with a measured rest-frame EW=3.0$\pm0.3 \AA$ and FWHM=8.8$\pm 1.7\AA$ ($\sim1610\pm300$ \kms).
The broad shape of the \heii\ emission line observed in the {\tt Sunburst} cluster is asymmetric and resembles a typical P-Cygni profile. 
The blue end of the emission line drops steeply, while the red end drops more gradually. 
The P-Cygni profile of \heii\ line is consistent with the profile predicted by models and synthetic spectra \citep[see][]{martins2022, Grafener2015}.
Broad \heii\ emission observed in galaxies is usually related to a non-nebular origin and is commonly associated with Wolf-Rayet (WR) stars \citep[e.g.,][]{Schaerer1999, Brinchmann2008, Grafener2015, Leitherer2018, Senchyna2021}, although the failure of the synthesis models to reproduce some of the strong \heii\ lines might be related to missing ingredients in stellar evolution models \citep[see, e.g.,][]{Leitherer2018}.
However, far-UV spectroscopic investigation of $\sim$57 individual stars located within the R136 star cluster revealed that massive stars with M>100 \msun\ are crucial for producing the \heii\ emission line \citep{Grafener2015, Crowther2016}.
On the other hand, \cite{martins2022} have shown that \heii\ can be produced in significant amount only when stellar winds are stronger than in normal O stars \citep[see also][]{Grafener2015}. 
VMS develop these strong winds because they lie close to the Eddington limit \citep{VG11,bestenlehner20,Graefener2021}. 
At the same time, these winds peel off the external layers of the stars and expose the products of hydrogen burning to the surface through the CNO cycle. 
This results in a strong nitrogen (and helium) enrichment that boosts the strength of \nivblue\ and \niv. 
This typically occurs after $\sim$1.5~Myr.
The two emission lines are both detected in the spectrum of the {\tt Sunburst} cluster at S/N~$>15$ (Figure \ref{1Dspec}), with EW=0.2$\AA$ and FWHM=2.9$\AA$ for \nivblue\ and EW=0.15$\AA$ and FWHM~$\sim2\AA$ for \niv. 
The helium enrichment also contributes to the strength of \heii. 
The same effects (strong winds combined with surface chemical enrichment) occur in normal evolved massive stars when they are seen as WR stars. 
The key difference to VMS is that helium enrichment takes place only after the main sequence  ($\ga 4$ Myr), while the same process takes place at younger ages in VMS.  
Furthermore, VMS are  more luminous than normal WR stars, and they accordingly contribute more strongly to integrated light.

The nebular \ha\ equivalent width provides constraints on the cluster age and hence on whether the \heii\ line is primarily due to WR stars or to VMS.
According to the BPASS models and results from \cite{Eldridge2012} (their Fig. 3), normal and WR stars produce EW$_{\rm H\alpha} <$ 1\AA\ for  $\la 3$Myr. 
The X-shooter spectrum reveals a prominent \ha\ line and no continuum detection, which very conservatively implies an equivalent width larger than 200\AA\ rest frame at $1\sigma$. 
However, when we assume for \ha\ the same continuum level as observed at $\lambda \sim 5000$\AA\ rest frame in the photometric spectral energy distribution (SED) by \citet{vanz22_CFE}, this limit increases to $\sim$ 840\AA. 
This value would be still a lower limit, even in the case of leaking ionizing photons.
After correcting the \ha\ flux for the fraction of escaping LyC photons (\ha$/(1-f_{\rm esc}^{\rm abs}$)), the resulting EW increases to EW$_{
\rm H\alpha} \sim$ 1231\AA.
We adopted $f_{\rm esc}^{\rm abs}$ values from \citet{rivera19}, where the corresponding absolute escape fraction of LyC photons along the line of sight is $f_{\rm esc}^{\rm abs}$ = 32$^{+2} _{-4}$\% . This large \ha\ equivalent width is consistent with a star-forming burst younger than $\sim 3$ Myr \citep[e.g.,][]{leitherer14}.
Furthermore, as discussed in \cite{chisholm19}, the \nv\ stellar wind profile predominantly depends on the stellar age, while variations due to different metallicity are negligible, and it is related to the young stellar populations ($<$ 5 Myr).
From the comparison of the observed \nv\ with the models Figure \ref{obs_VS_models_150} and \ref{obs_VS_models}, we additionally demonstrate that the age of the cluster is $< 3$ Myr.
From our age analysis, we can conclude that the properties of both \nv\ and \ha\ fit a $< 3$ Myr age of the stellar cluster well, which requires other sources than WR stars to explain the observed strong \heii\ EW=3.0$\pm0.3 \AA$.
Therefore, these results strongly suggest that VMS produce the spectral ultraviolet features we observe in this young massive star cluster.
Moreover, \cite{Wofford2014} and \cite{smith16} have argued that \ov\ in the integrated light of the clusters is a key feature of VMS as well. 
This line is not seen in the {\tt Sunburst} cluster; see Figure \ref{1Dspec}.
As demonstrated by \cite{martins2022}, this is not incompatible with the presence of VMS because \ov\ disappears as VMS evolve to a lower effective temperature. 
In their Fig.~4, no sign of \ov\ exists after $\sim$1~Myr.  
Together with \nivblue, this places a rather tight constraint on the cluster age.

\subsection{Comparing observations with models}
\label{sec:obs vs models}

To investigate the rest-frame UV spectrum of the cluster, we created an integrated VMS model following \cite{martins2022} that includes normal-mass stars (0.1-100 \msun) with different VMS masses (150 \msun\ and 200 \msun).
It is important to note that our models rely on the predictions of \cite{martins2022}, which are based on the empirical mass-loss prescriptions of \cite{Graefener2021}.
Different mass-loss prescriptions can affect the evolution of VMS, as shown by \cite{Sabhahit2022}, for instance.
The HI ionizing budget mostly depends on the effective temperature ($\rm T_{eff}$) and less on an exact wind prescription \citep[e.g.,][]{martins2022}.
Furthermore, because winds can strongly affect stellar evolution and thus $\rm T_{eff}$, the estimate of the contribution of VMS to the ionizing budget we presented in Section \ref{sec:LyC budget} is model dependent. 
However, we stress that our models are able to reproduce the integrated spectroscopic properties of R136 (Martins et al. in preparation) and are thus appropriate.

We used the spectral energy distribution (SEDs) of the BPASS \citep{eldridge2017,eldridge2018} v2.2.1 single-star population synthesis model. 
The model has an upper mass limit of 100 \msun\ with a Salpeter IMF, a metallicity of Z=0.006 (where 0.02 corresponds to solar metallicity), and an instantaneous star formation history, with a burst of mass $10^6$ \msun. 
The adopted metallicity of the model is closest to our measured value based on the N2 index \citep{Marino2013}, which is $\simeq 0.4 Z_{\odot}$ and is consistent with the estimate provided by \citet{Mainali2022} (see also \citealt{chisholm19}).  

We extrapolated the Salpeter IMF to a 225 \msun\ upper-mass limit within a few mass bins given by Equation \ref{eqBPASS} from the BPASS manual\footnote{https://flexiblelearning.auckland.ac.nz/bpass/9.html}. 
Equation \ref{eqBPASS} gives the number of massive stars in the mass range [$M_a$; $M_b$],
\begin{equation}
\begin{aligned}
    N(M_a;M_b) = C \times M_{1}^{\alpha_1} \int_{M_a}^{M_b} M^{\alpha_2} \,dM\
\label{eqBPASS}
\end{aligned}
.\end{equation}
Here, $C$ is a constant and has the value of $1.23 \times 10^5$ for an arbitrary burst mass of $10^6$ \msun. 
Moreover, $M_1$ = 0.5 \msun\, $\alpha_1$ = -1.3, and $\alpha_2$ = -2.35. 
The mass bins were selected in a way to add the SEDs of appropriate numbers of single VMS stars, which are available for discrete sets of VMS with masses including 150 \msun\ and 200 \msun.
In this manner, we computed SEDs including VMS with IMFs extending up to 175 and 225 \msun, respectively, following \cite{martins2022}.
In the mass bin from 100 \msun\ to 175 \msun, the extrapolation of the Salpeter IMF using Equation \ref{eqBPASS} gives $\sim237$  VMS, and for the mass bin from 175 \msun\ to 225 \msun, it provides us with $\sim60$ VMS. 
We added SEDs of 150 \msun VMS as an average of the VMS population in the mass bin 100 $\msun \le \rm M \le 175 \msun$, and we did the same by adding SEDs of 200 \msun\ VMS for the mass bin 175 $\msun \le \rm M \le 225 \msun$.
Finally, these numbers have to be rescaled to the appropriate cluster mass, which is $\sim 10^7$ \msun\ \citep{vanz22_CFE}.

The YMC shows significant \nivblue \ emission, as shown in Figure~\ref{1Dspec}.
From the VMS models and synthetic spectra, \nivblue \ emission only appears after 1.5 Myr of VMS evolution \citep[see][]{martins2022}, and VMS last approximately until 2.5 Myr. 
Based on this, we created the SEDs of integrated VMS models at 1.5 Myr, 2 Myr, and 2.5 Myr. 
We normalized the spectrum of the cluster and the models by fitting a UV power law by using the spectral windows provided by \cite{rix2004}.

\begin{figure*}
        \centering
        \includegraphics[width=\linewidth]{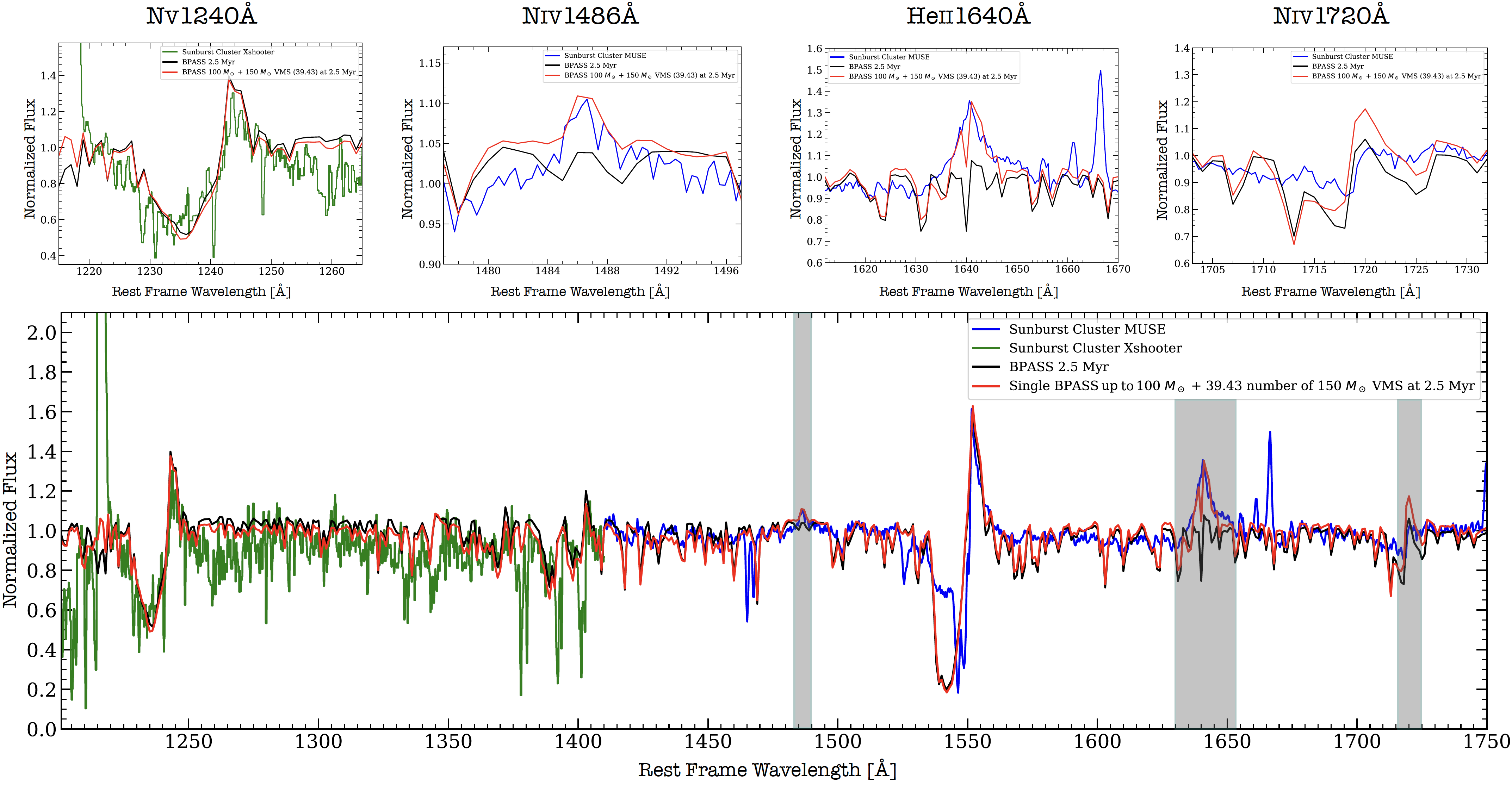}
        \caption{ MUSE ultraviolet spectra of the young star cluster (blue, bottom panel) with the X-shooter spectrum (green). The shaded gray regions show specific UV features that are closely associated with the VMS (\nivblue, \heii, and \niv), and some of them show a P-Cygni profile, as is characteristic of young and massive stars. Furthermore, the black line shows the single BPASS model including only normal stars at 2.5 Myr, and the red line shows the BPASS single-star model augmented by VMS with masses up to 150 $\msun$.
        The upper panels show the zoom in VMS characteristic features compared with the models and the \nv\ P-Cygni line, which is characteristic of very young stellar populations.}
        \label{obs_VS_models_150}
\end{figure*}

\begin{figure*}
        \centering
        \includegraphics[width=\linewidth]{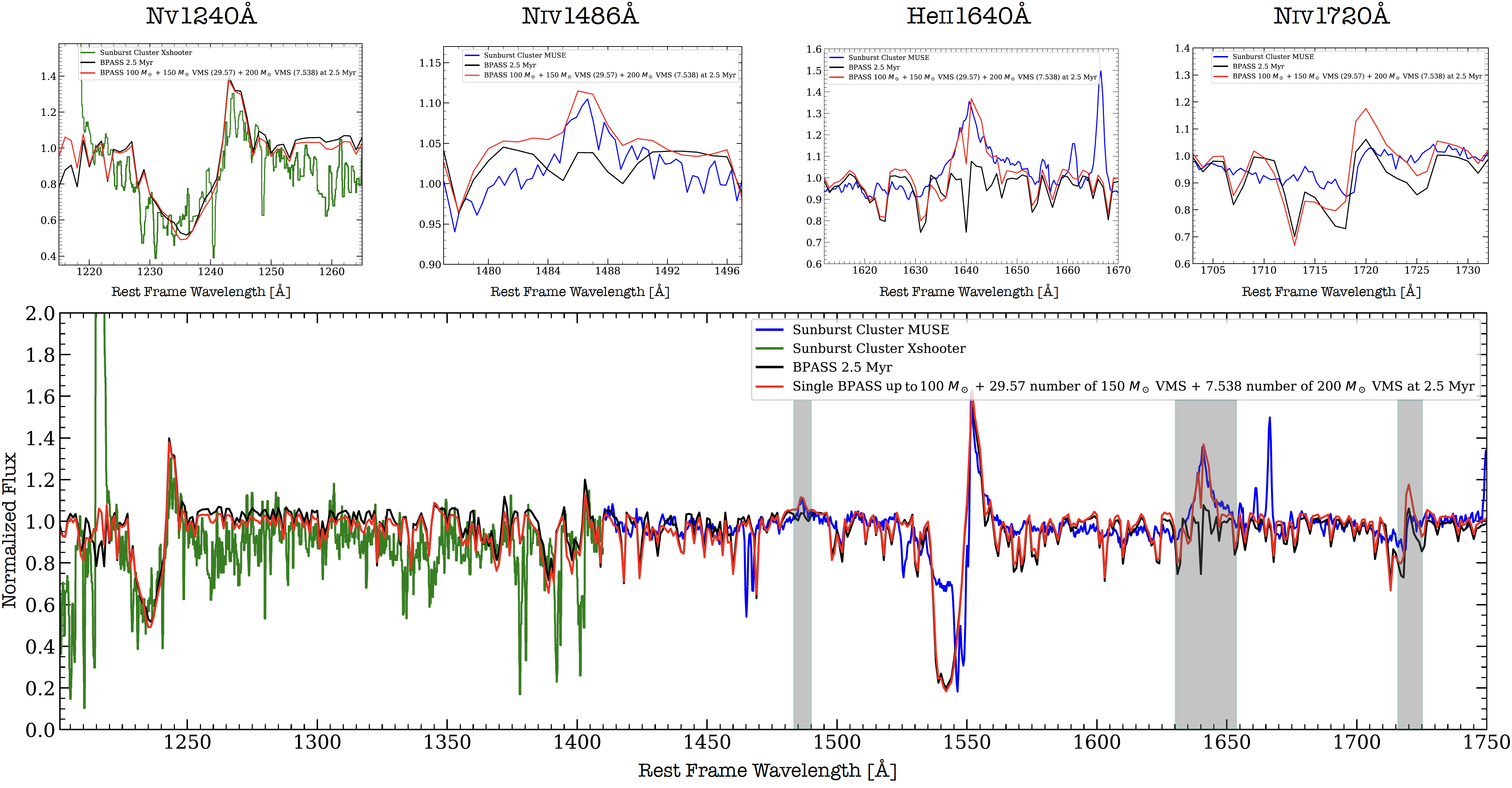}
        \caption{ All symbols are the same as in Figure \ref{obs_VS_models_150}, except for the red lines,  which show the BPASS single-star model augmented by VMS with masses up to 200 \msun.
        }
        \label{obs_VS_models}
\end{figure*}

We directly compared the cluster spectrum with the two VMS models at three different ages. 
The comparison shows that VMS are clearly needed to reproduce the observations (see Figures~\ref{obs_VS_models_150}, \ref{obs_VS_models}, and \ref{appendix_VMS_full_contribution}). 
However, the \heii \ and \niv \ lines in the models appear stronger than observed even at an age of 1.5 Myr and with a maximum mass of 175 \msun. 
To match the observed \heii \ and \niv \ profiles, we therefore reduced the VMS contribution by decreasing their numbers. 
When the VMS contribution is reduced by a factor of 6 in the VMS model, which includes only 150 \msun\ VMS at 2.5 Myr (Fig.~\ref{obs_VS_models_150}), the results agree well. 
Alternatively, a similar match is also found by reducing the VMS contribution by a factor of 8 in models that also incliude the 200 \msun\ VMS (Fig.~\ref{obs_VS_models}). 
\textcolor{blue}{For a cluster mass of $10^7$ \msun\ , this implies a total of $\sim 370-400$ VMS for both cases.}
In short, the observations are compatible with an IMF extending up to $\sim$ 175 or 225 \msun, but with an IMF slope steeper than for a Salpeter IMF ($\alpha_2 <-2.35$) for $M > 100$ \msun.

\subsection{VMS contribution to the Lyman continuum budget of the young stellar cluster}
\label{sec:LyC budget}

The contribution of the VMS to the total LyC emission can be determined in a straightforward way from the theoretical SEDs by comparing models
with and without VMS by integrating the theoretical SEDs in the defined LyC range. From the favored model including VMS, but with a steep IMF slope, and a model excluding VMS, we find that VMS contribute $\sim$15\% of the LyC
radiation that is emitted over the range $\sim$730--900\AA\ rest frame (covered by HST F275W filter). Similar percentages are also obtained by integrating over the total LyC.
For comparison, we note that this percentage is somewhat lower than the fraction of LyC photons produced by VMS in R136, which is found to be 25-40\% \citep{Doran2013,Crowther2010}. This is consistent with our inference of fewer VMS than in a standard IMF.
Since LyC photons are mainly produced by O-type and more massive stars, $\sim$15\% of the LyC production is generated from only $\sim$1\%\ of the stars that are capable of producing LyC photons.

\section{Spatial segregation of the Lyman continuum radiation}
\label{sec:morpho}

\subsection{Morphological analysis of 5.1l YMC}

We now address the morphological properties of image 5.1l, which is the most strongly magnified of the multiple images of the star cluster ($\mu_{tot} \simeq 76$; \citealt{Pignataro2021}).
Image 5.1l is the brightest image and was detected with a high S/N in the F275W (S/N $\sim90$) and F555W (S/N $\gg$100), allowing us to investigate and compare the morphology in these two spectral regions: the emitting LyC ($\lambda < 900$\AA in HST F275W band) and the nonionizing radiation at 1700\AA\ (HST F555W band). We followed two approaches: (1) We ran simulations injecting the sources in the F275W band, and (2) we analyzed the curve of growth of the resulting images.

\begin{figure}
        \centering
        \includegraphics[width=0.5\textwidth]{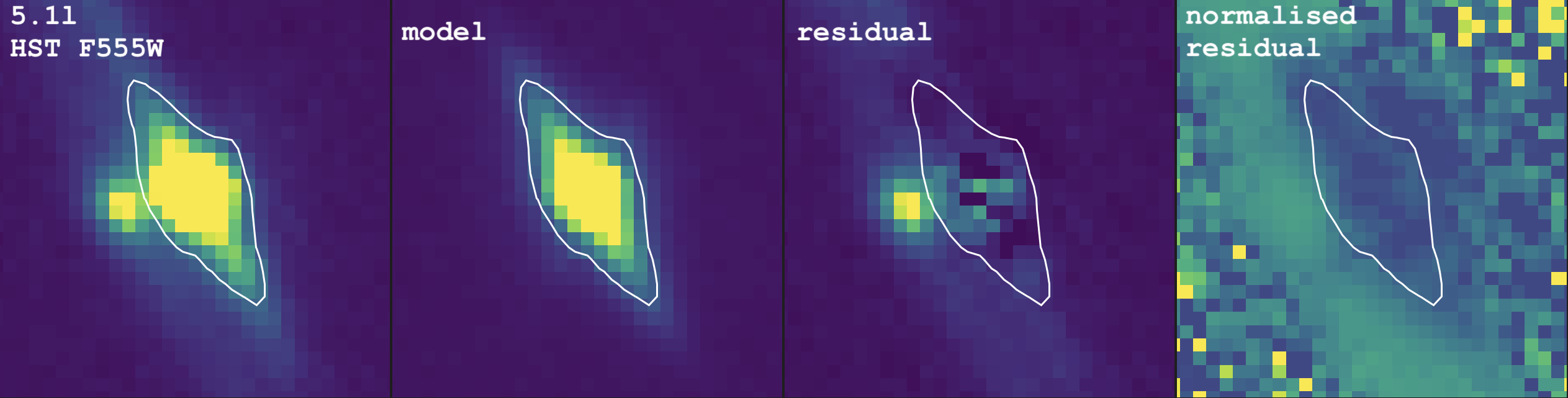}
        \caption{Results from Galfit modeling after using a two-component fit.
        From the left, the first panel shows source 5.1l in the F555W band (UV1700\AA). The second panel shows the model from Galfit. The third panel shows the residual, and the fourth panel is the normalized residual produced after dividing the residual with the original image. The white contour encloses the region we used to check the quality of the produced model.}  
        \label{galfit_2comp_model}
\end{figure}

Figure~\ref{galfit_2comp_model} shows the F555W image of 5.1l, in which the elongation is clearly visible in the direction of the tangential stretch produced by gravitational lensing. 
As discussed in \citet{Pignataro2021} (see also \citealt{vanz22_CFE}), the tangential amplification largely dominates along the arc ($\mu_{tang} \simeq 57$).
We perform a relative comparison between images to ensure that the results do not depend on the magnification values.

\begin{figure}
        \centering
        \includegraphics[width=0.5\textwidth]{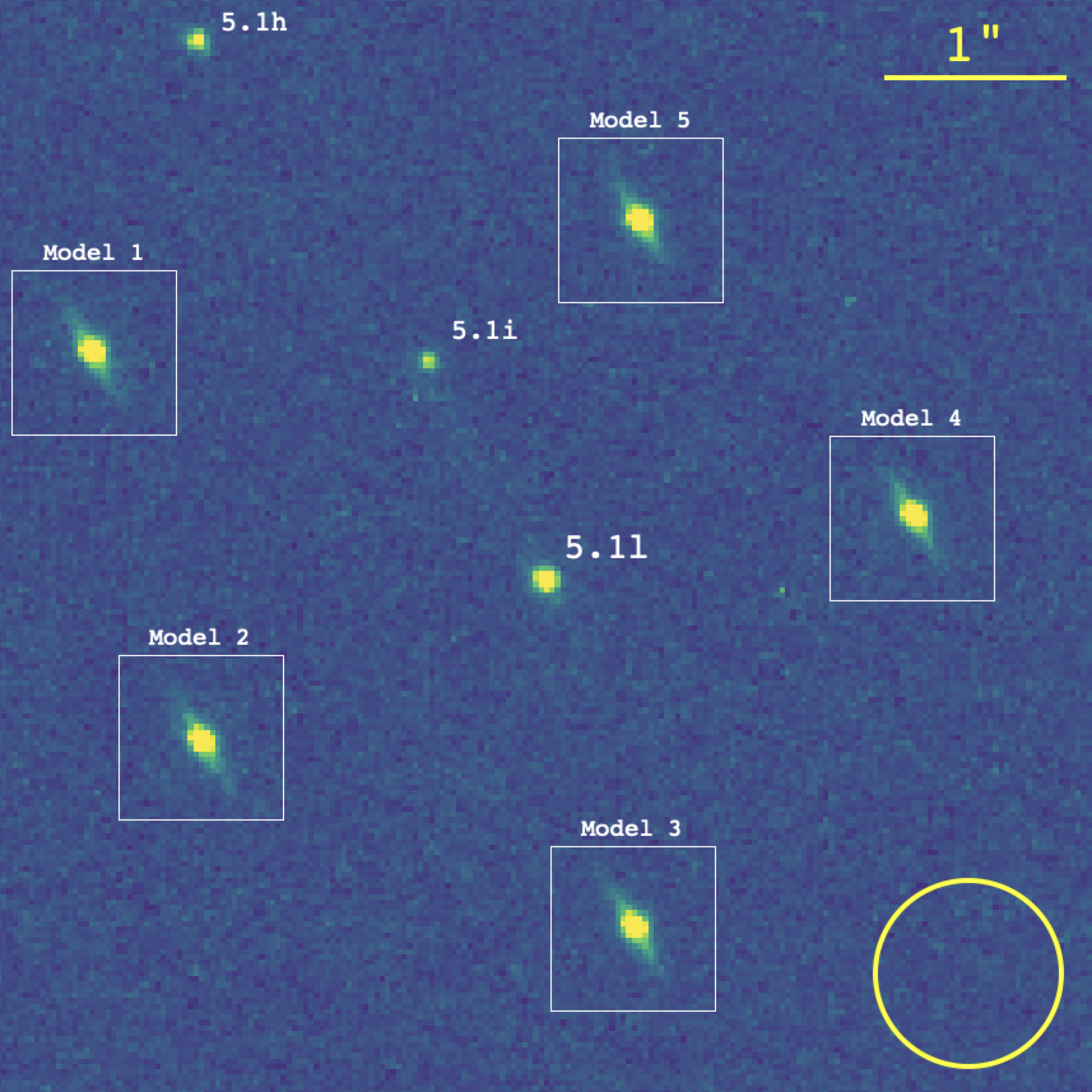}
        \caption{HST F275W image showing source 5.1l leaking LyC in its center. The other two multiple images of the same source are labeled 5.1i and 5.1h. In the region not contaminated by other sources, five models are injected around source 5.11 (see text for more details). They are enclosed in the white squares. In the bottom right corner, we show the largest (diameter of 34pix) aperture size used to construct CoGs.}
        \label{sim_n05}
\end{figure}

As a first step, we computed a realistic model of image 5.1l in the HST F555W image using {\tt Galfit} \citep[][]{Peng_2010}. 
The point spread function (PSF) was extracted by combining unsaturated stars in the field of view. While the fit with a single component did not produce acceptable residuals (higher than 20\%), we reproduce the light profile of the object quite well by combining two components: a core with a Gaussian light profile and an effective radius (R$_{\rm eff}$) smaller than 0.5 pixels (in practice nearly unresolved), and an extended component with R$_{\rm eff}=6$ pixels and a S\'ersic index n=1 (similar results are also obtained with n=0.5). 
The combination of the two components produces an optimal shape that leaves normalized residuals lower than 10\% (see Figure~\ref{galfit_2comp_model}). 
It is worth investigating whether this resolved shape (sampled at 1700\AA) is recovered when it is placed in the F275W LyC image.
For this check, we injected mock images of source 5.1l into the F275W image in five different positions around source 5.1l that were not contaminated by the flux from other sources. 
These images were produced from the aforementioned two-component {\tt Galfit} model constructed at 1700\AA\ (F555W), but now accounting for the F275W PSF (i.e., convolved by the F275W PSF) to allow for a proper comparison with the LyC source 5.1l (observed in the HST F275W band).
These images were added to F275W after each of them was rescaled to the observed peak value of the  LyC source 5.1l. 
This step was performed with the {\tt IRAF} \citep{Tody1986} tasks {\tt IMARITH} and {\tt IMCOPY}.
Figure~\ref{sim_n05} shows the results, in which all the injected images show a spatially resolved morphology along the tangential magnification. 
Conversely, the observed LyC image (of source 5.1l) appears nucleated, suggesting that the emitting LyC region is smaller than the region at 1700\AA.

\begin{figure}
        \centering
        \includegraphics[width=\linewidth]{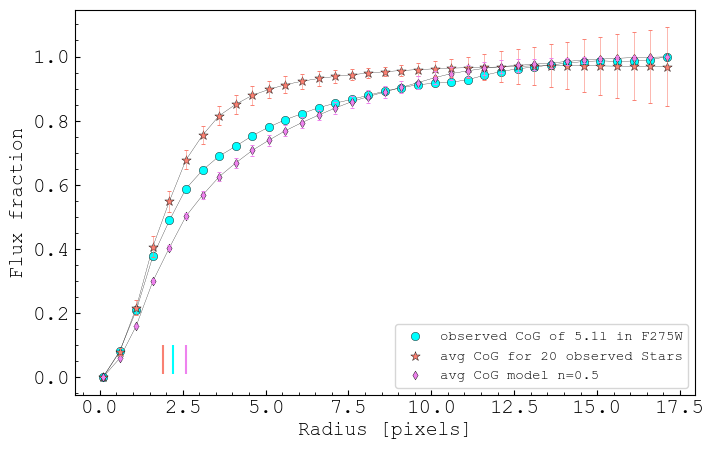}
        \caption{Three curves of growth normalized to 1. The cyan CoG corresponds to source 5.1l leaking LyC that is located in the {\tt Sunburst} arc (F275W band). The violet CoG is the PSF-convolved best-fit model of source 5.1l observed in F555W, constructed by averaging 5 CoGs (see text for more details). The orange growth curve is used for comparison and was constructed by averaging 20 single CoGs from randomly selected stars. In both cases (cyan and violet), the error bars are $1\sigma$. Vertical lines in the bottom left part of the panel mark the pixel radii within which 50\% of the light is enclosed. The colors corresponds to the CoG colors. }
        \label{cog}
\end{figure}

To quantify this result, we calculated the curve of growth (CoG) of the images shown in Figure~\ref{sim_n05}.
The flux was then measured in the F275W band in 34 circular apertures.
The smallest aperture had a radius of 0.1 pixel.
Intermediate apertures were drawn with increasing radii, with a step of 0.5 pix, up to the largest aperture, which has a radius of 34 pixels.
As a reference point-like source, we constructed the mean CoG from a selected sample of 20 unsaturated and uncontaminated stars.
The resulting CoG is shown in Figure~\ref{cog}, where the y-axis reports the fraction of the flux that is enclosed in the corresponding radius in pixels (x-axis).

The same procedure was applied to the LyC-emitting source 5.1l,
while another CoG was constructed by averaging the five CoG of the injected models that resemble the morphology at 1700\AA.
Figure~\ref{cog} compares all the CoGs after normalizing them to the saturation value at the largest radius.
The first result from this test is that the CoG of the observed 5.1l LyC source clearly deviates from the behavior of a point-like source (stars). 
This was not explored before and suggests that in the most magnified image of the star cluster, the LyC appears to be spatially resolved. 
{\it \textup{This is the first evidence of resolved stellar LyC emission at cosmological distance}.} 
Second, this barely resolved LyC emission appears to be more nucleated than the emission at 1700\AA. Consequently, the sources of ionizing radiation appear to be located in the central part of the cluster.
These curves show that 50\% of the stellar flux is enclosed within a radius of $\sim$1.9 pixel, while for source 5.1l, the flux lies within $\sim$2.2 pixels. 
Additionally, we performed a Kolmogorov-Smirnov two-sample test (KS-test) to determine whether the CoGs derived from the stars and from source 5.1l follow the same distribution (null hypothesis).
For this purpose, we used the statistical function {\tt ks\_2samp} from {\tt scipy.stats}. 
After comparing the average CoG of the stars with cyan and violet CoGs, the KS-test gives $p << 0.05$.
This means that the null hypothesis is not satisfied with the LyC profile of source 5.1l, and it deviates from the CoG of a point-like source.
Furthermore, we also find that the half-light size of 5.1l at 1700\AA\ is larger than the ionizing region, $\sim$2.6 pixels compared to the $\sim$2.2 pixels.
When we correct these radii for the instrumental resolution (given by the stars), we obtain an effective radius for source 5.1l at 1700\AA\ of\ $\simeq 7.8\pm1.4$~pc after delensing\footnote{Adopting the pixel scale of $0.03''$/pixel, 8200 pc per arcsecond at z=2.37 and $\mu_{tang} \simeq 57$, R$_{\rm eff}=0.03*8200*((2.6^{2} - 1.9^{2})^{0.5})/57$, adopting the same uncertainty on $\mu_{tang}$ as reported by \citet{vanz22_CFE}.}. 
This agrees with \cite{vanz_sunburst},
while the LyC image (5.1l) has a smaller radius, R$_{\rm eff} \simeq 4.7\pm1.5$ pc.  
The LyC emission is therefore more compact than the nonionizing UV continuum, which we interpret as a spatial segregation of the most massive stars.

\subsection{Segregation of (very) massive stars?}

The segregation of massive stars is observed in young stellar clusters
\citep[][]{Harayama2008, Habibi2013}, and its origin is still an open question. 
Two scenarios for mass segregation within young star clusters have been proposed.  The first scenario suggests that mass segregation is primordial \citep[e.g.,][]{Dib2010}.
The second scenario that can explain mass segregation requires dynamical evolution.
In this case, massive stars migrate to the central parts of the cluster as a consequence of two-body encounters with less massive stars \citep[e.g.,][]{Binney1987}.

Mass segregation has so far been investigated and reported in several stellar clusters located in the Milky Way \citep[e.g., NGC3603;][]{Harayama2008} or in the Local Universe \citep[e.g., R136;][]{Khorrami2021}. 
However, at cosmological distances as in the case of source 5.1 YMC presented in this work, 
the analysis described in the previous section suggests that the regions emitting ionizing and nonionizing radiation have different sizes and that the bulk of LyC radiation arises from the core of the star cluster.
This implies that the only producers of LyC radiation, that is, hot massive stars, 
are segregated in the central parts of the cluster and are surrounded by a more diffuse distribution of less massive, non-LyC emitting stars.

Dynamical relaxation occurs on timescales comparable with the relaxation time $t_{relax}$. When $t_{relax}$ exceeds the age of the cluster, the most likely origin for segregation is therefore primordial. The relaxation time in the center of the cluster can be estimated as $t_{relax}\backsimeq(0.065 \times \sigma^{3})/(G^{2} \rm m \rm \rho \rm ln\Lambda$) \citep{Spitzer1987}, where $\rm m$ is the average stellar mass in the cluster (the mass of the cluster divided by the total number of stars), $\sigma$ is the central three-dimensional velocity dispersion, $\rho$ is the central mass density, $G$ is the gravitational constant, and $\ln\Lambda$ is the Coulomb logarithm, a number typically between 6 and 15. To compute the central relaxation time of source 5.1l YMC, we assumed that its stellar distribution is well approximated by that of an isotropic self-gravitating \cite{Plummer1911} model. In a Plummer model, the central velocity dispersion is $\sigma=\sqrt{\rm GM/(6r_{0.5})}$.
We adopted a mass of $\rm M = 10^7\msun$ , as estimated by \cite{vanz22_CFE}, 
and we assumed that in the cluster, light traces mass, so that its half-mass radius is equal to the half-light radius $r_{0.5}$=8pc, as measured in F555W. The resulting central velocity dispersion is $\simeq$30km/s. This value agrees with other extremely massive star clusters reported in the literature \citep{bastian06, ostlin07}. The number of stars was calculated from equation 2.4 of \cite{eldridge2018} and was rescaled to the mass of source 5.11. The estimated number of stars in source 5.1l YMC is N$\sim9\times10^6$. Combined with the total cluster mass, this gives an average stellar mass of $m\sim1.1$M$_\odot$.
The central density in a Plummer model is $\rho = 3\rm M/(4\pi  r_{0.5}^3)$, and we adopted $\rm ln\Lambda=15$.
The resulting $t_{relax}\sim1.1$ Gyr clearly implies that stars did not have enough time to change their initial trajectory, which further indicates that the observed segregation is most likely of primordial nature. We also note that our estimate of the relaxation time is a lower limit: the cluster center in our model lies in the place in which the ratio of $\sigma^3$ and $\rho$ is lowest. The same measure performed at any larger distance from the cluster center would have given an even longer $t_{relax}$.

\section{Summary and conclusions}
\label{sec:Summ}

We have presented a detailed spectroscopic and morphological analysis of the massive and young stellar cluster hosted in the lensed galaxy called {\tt Sunburst} galaxy at z=2.37, for which LyC emission was also confirmed in the literature.
We used results from recent stellar evolutions and atmosphere models including VMS \citep{martins2022} to conduct extensive comparisons with high spectral resolution observations performed with VLT/MUSE and X-Shooter.
The main results of this work can be summarized as follows:
1) In the spectroscopic observations, the high S/N MUSE and X-shooter spectra reveal features of broad (and asymmetric) \heii\ emission with an EW~$\simeq3$\AA\ in the rest frame and a line width of 1610 \kms, and \nivblue\ with an EW~$\simeq 0.2$\AA\ in emission. In addition, the P-Cygni profile of \niv\ (along with NV and CIV) is also observed. 
All these features suggest the presence of very massive ($> 100$\msun) stars.  
The absence of \ov\ provides a lower age limit of 1 Myr.
On the other hand, the large \ha\ EW ($>1231$\AA\ after correcting for the escaping LyC radiation) indicates an age younger than $\sim$ 3 Myr.
These narrow age constraints strongly favor the existence of VMS over WR stars, implying that the strength of the \heii\ emission line is entirely due to VMS.
2) A comparison of the observations with the models revealed that the most plausible age of the star cluster is 2.5 Myr, and the estimated number is $\sim 370-400$ VMS for a cluster mass of $10^7$ \msun.
The observations are compatible with an IMF extending up to $\sim 175-225$ \msun, but with a steeper slope than the Salpeter IMF.
3) The fraction of LyC radiation emerging from the VMS component is not negligible. 
We estimate that in the 730\AA\ -- 900\AA\ range (probed by the HST/F275W band), about 370 -- 400 VMS (or roughly 1\% of the total population of O-type stars in the star cluster) account for 15\% of the escaping LyC photons. The remaining photons are mostly produced by the other less massive O-type stars.
4) Detailed morphological analysis of the most magnified image of the star cluster shows that the region emitting LyC is not point-like, with a light profile different from the average profile of stars present in the same field of view. This is the first evidence of a resolved LyC emission at any redshift. Remarkably, the physical scale of the LyC emitting region also appears to be smaller (with a significant K-S probability $p<<0.05$) than the nonionizing region (1700\AA), suggesting that massive O-type stars that cause the LyC radiation, and likely the VMS (significantly contributing to it), are segregated in the central part of the star cluster. After delensing the angular half-light radii, the LyC region appears barely resolved with R$_{\rm eff} \simeq 4.7 \pm 1.5$pc, while at 1700\AA,\ it is R$_{\rm eff}\simeq 7.8 \pm 1.4$ pc. The packaging of this large number of massive O-type stars per parsec cube in the central region, $\simeq 70$~pc$^{-3}$, is likely a element that allowed carving the ionizing channel and the development of a high-speed outflowing gas \citep[][]{rivera17, vanz22_CFE, Mainali2022}.

As a final remark, it is important to stress that we investigated this remarkable object and its exotic properties based on gravitational lensing. 
Otherwise, this investigation would be impossible with current instrumentation in blank fields at cosmological distances. 
This once again demonstrates the potential and importance of gravitationally lensed systems for investigations of the most exotic phenomena in the Universe.


\label{sum}

%

\begin{acknowledgements}
We acknowledge financial support through grants PRIN-MIUR 2017WSCC32, 2020SKSTHZ and the INAF GO Grant 2022 “The revolution is around the corner: JWST will probe globular cluster precursors and Population III stellar clusters at cosmic dawn” (PI Vanzella). FC and RP acknowledge funding from PRIN INAF 1.05.01.85.01.

\end{acknowledgements}

%
%

\bibliographystyle{aa}
\bibliography{bibliography}

\begin{appendix} 

\section{Initial models}
\label{ha}

\begin{figure*}
        \centering
        \includegraphics[width=\linewidth]{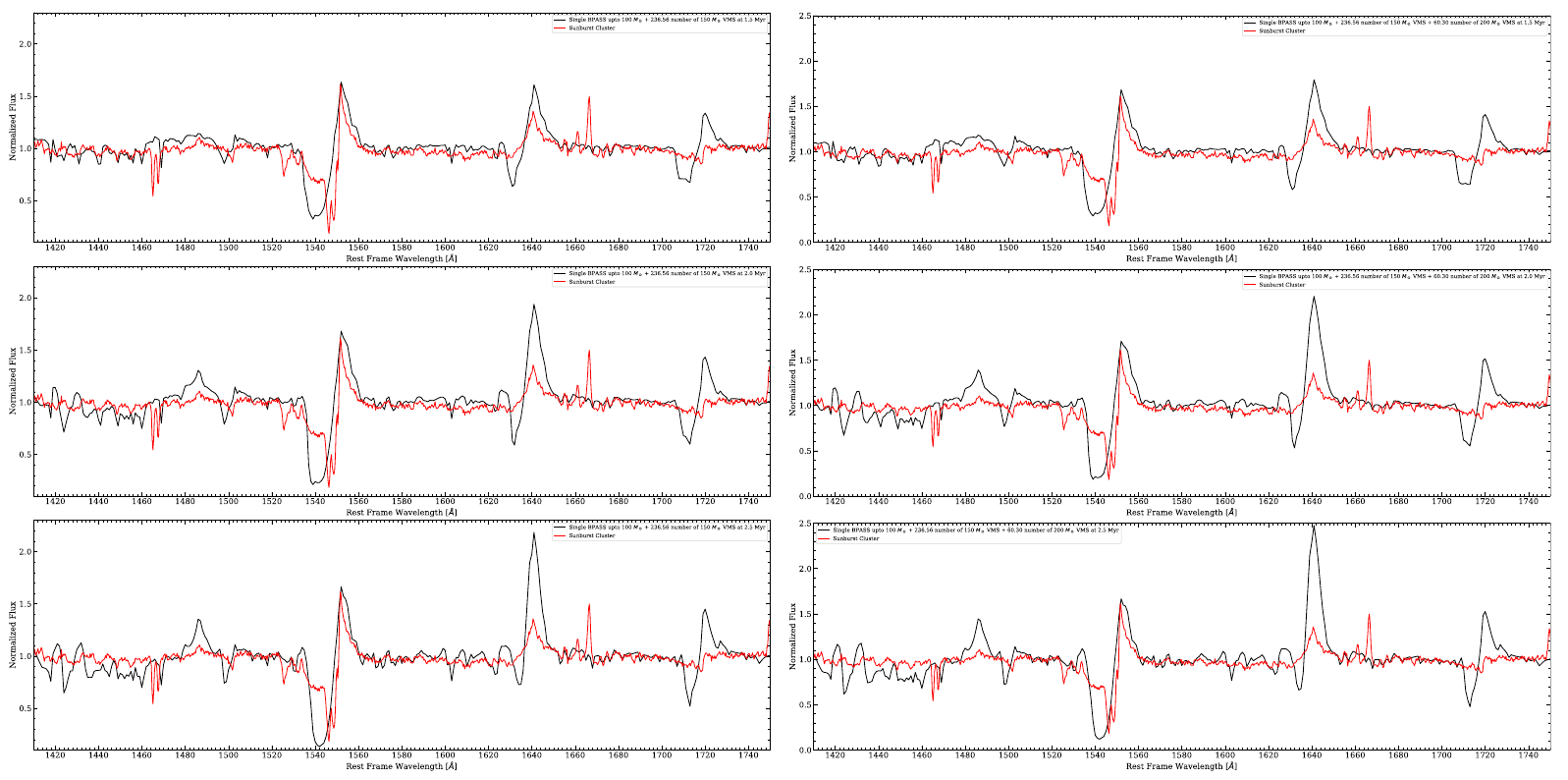}
        \caption{Comparison of the observations (MUSE spectrum, red line) with BPASS models (black line). The left column shows three BPASS models with an IMF up to 100 \msun, ages of 1.5 Myr, 2 Myr, and 2.5 Myr and an added number of 236.56 150 \msun\ VMS. The right column shows a BPASS model with an IMF up to 100 \msun\ at same ages (as shown in previous column), but with an added number of 236.56 and 60.30 VMS of 150 \msun\ and 200 \msun.}
        \label{appendix_VMS_full_contribution}
\end{figure*}

As described in Section \ref{sec:obs vs models}, we compared the X-shooter and MUSE spectroscopic observations with the BPASS models at different ages (1.5 Myr, 2 Myr, and 2.5 Myr).
We narrowed our models to the mentioned age range because the predicted age of the cluster exceeds 1.5 Myr (inferred from the \nivblue\ emission line) and the lifetime of the VMS is about 2.5 Myr \citep{martins2022}.
We started with the BPASS that has an upper mass limit of 100 \msun\ and added 236.56 stars in the 100 - 175 \msun\ mass range and 60.30 stars in the mass range 175 -- 225 \msun.
The resulting models (at different ages) are shown in Fig. \ref{appendix_VMS_full_contribution}. All models produce significantly strong spectroscopic features that are characteristic of VMS.
To better match models with observations, we decreased the numbers of the VMS, and the final results are presented in Section \ref{sec:VMS} (Figs. \ref{obs_VS_models_150} and \ref{obs_VS_models}).

\end{appendix}

\end{document}